\documentclass[lettersize,journal]{IEEEtran}
\usepackage{amsmath,graphicx,booktabs,tabularx,cite,url,tcolorbox}
\definecolor{TakeawayInk}{HTML}{19577D}
\definecolor{TakeawayTint}{HTML}{EAF3F9}
\newtcolorbox[auto counter]{takeaway}{
  colback=TakeawayTint,colframe=TakeawayInk,boxrule=0pt,leftrule=1.7pt,
  sharp corners,boxsep=0pt,left=5pt,right=5pt,top=3pt,bottom=3pt,
  before skip=4pt,after skip=4pt,
  fontupper=\fontsize{9}{10.5}\selectfont,
  before upper={\textcolor{TakeawayInk}{\textbf{Takeaway \thetcbcounter.}}\enspace}
}
\usepackage[hidelinks]{hyperref}
\graphicspath{{figures/}}
\IEEEaftertitletext{\vspace{-3em}}

\begin{document}
\title{When Is High-Bandwidth Flash Cost-Effective for LLM Inference?}
\author{Zhiyu An, Yiping Li, and Wan Du\\

\thanks{Authors are with the Department of Computer Science and Engineering,
University of California, Merced, Merced, CA, USA.
(e-mail: zan7@ucmerced.edu, wdu3@ucmerced.edu).}%
}
\maketitle
\begin{abstract}
Can high-bandwidth flash (HBF) make LLM inference cheaper than adding GPUs? 
By storing weights closer to the GPU than SSD, HBF may reduce the number of GPUs needed for model capacity. 
However, this integration is only justified if it meets service targets and its cost is recovered. We simulate this integration using public component data across nine dense and mixture-of-experts (MoE) models, five workloads, two GPU platforms, and 1--100 GPUs. At fixed demand, HBF reduces the minimum GPU count to meet a service target in 98 of 468 comparisons; majority of them are paused and short sessions, 63 remain cost-effective at half a GPU price per HBF stack. For single short-chat users, HBF enables large MoEs, e.g. DeepSeek V3 and Kimi K2, to meet a relaxed service target on one RTX~4090 GPU, where conventional serving fails. At fixed GPU count, HBF improves throughput in 48\% of matched comparisons below 16 GPUs and 13\% at larger counts. Even at 100 RTX~4090-scale GPUs, cascaded HBF raises offline throughput by 16.4--26.5$\times$ for three large MoEs. We found savings and throughput gains arise in different scenarios.

\end{abstract}

\section{Introduction}
\IEEEPARstart{W}{hy} buy more GPUs if the bottleneck is memory capacity rather than compute? Large LLM weights can force deployments to distribute a model across GPUs even when the workload needs little computation. High-bandwidth flash (HBF) could hold infrequently accessed weights close to the GPU, reducing the GPU count or enabling a model to run on one GPU. But for HBFs to be cost-effective, flash reads must meet latency target; replacing near memory can hurt performance, and HBF stacks add cost.

Existing studies highlight this tradeoff. Son \emph{et al.} examine weight/KV placement and flash bandwidth and endurance~\cite{son}.
Li \emph{et al.} show slowdowns from near-memory displacement and transient KV writes~\cite{kv}. 
\textit{Unlike their setup of HBF as a primary replacement against HBM-only systems, we study HBF alongside retained near memory and conventional DRAM/SSD offloading, quantifying when its GPU savings justify the added hardware cost.}
H$^3$ and FlashAccel study hybrid hierarchies and weight streaming~\cite{h3,flashaccel}. 
\textit{Here we isolate the case for read-mostly HBF: active KV stays near the GPU, while cold weights compete with conventional host-DRAM/SSD offload~\cite{flexgen,suram}, which HBF could replace.}

\begin{figure}[!t]
\centering
\includegraphics[width=\columnwidth]{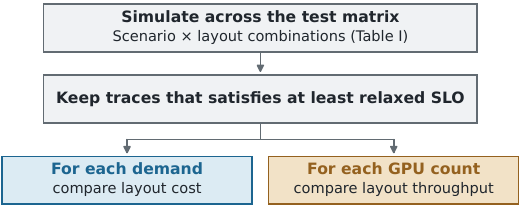}
\caption{Comparison workflow for interactive workloads. Each branch compares layouts at the selected SLO.}
\label{fig:workflow}
\vspace{-1em}
\end{figure}

\begin{figure}[!t]
\centering
\includegraphics[width=\columnwidth,trim=0 13bp 0 0,clip]{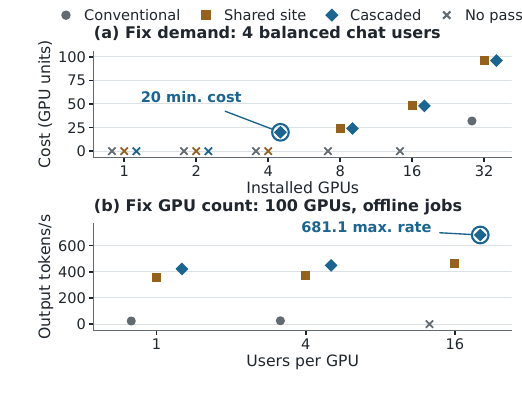}
\caption{RTX~4090-scale examples: (a) Qwen3-235B minimum cost at one GPU cost/stack: 4 GPUs, 16 stacks; above 32 GPUs, cost $\geq64$. (b) Kimi K2 maximum rate within one hour. Crosses: no pass; rings: winners.}
\label{fig:decisions}
\vspace{-2em}
\end{figure}

We therefore ask two questions. \emph{For a given demand}, can HBF meet the service-level objective (SLO) with fewer GPUs, and do those savings pay for the stacks? \emph{For an existing GPU installation}, can HBF serve more requests while meeting the SLO? We simulate nine dense/MoE models, five workloads, two platforms, and 1--100 GPUs (Table~\ref{tab:matrix}). Figures~\ref{fig:workflow} and~\ref{fig:decisions} summarize the comparisons. These comparisons separate four outcomes that are often conflated: fewer GPUs, newly feasible models, lower purchase cost, and higher throughput.

\begin{table*}[!t]
\caption{Scenario axes and layout test matrix}
\label{tab:matrix}
\centering
\footnotesize
\setlength{\tabcolsep}{3pt}
\begin{tabularx}{\textwidth}{@{}p{.111\textwidth}>{\raggedright\arraybackslash}X
    @{\hspace{15pt}}p{.113\textwidth}>{\raggedright\arraybackslash}X@{}}
\toprule
\multicolumn{2}{@{}l}{\textbf{Scenario axes}} &
\multicolumn{2}{@{}l@{}}{\textbf{Layout test matrix}} \\
\cmidrule(r{5pt}){1-2}\cmidrule{3-4}
\textbf{Models} &
\textit{Dense} Qwen3-8B/32B, Llama 3.1-70B/405B\newline
\textit{MoE} Qwen3-30B-A3B, Qwen3-235B-A22B, DeepSeek V3, Kimi K2, Muse Spark Open &
\textbf{Memory} &
Near-memory sites + HBF stacks per GPU\newline
Conventional $8+0$, shared-site $6+2$ or $4+4$, cascaded $8+2$ or $8+4$ \\
\addlinespace[2pt]
\textbf{Platforms} & RTX~4090-scale GDDR, H200-scale HBM &
\textbf{Weight path} & Resident, GPU staging, or CPU experts (conventional MoE control) \\
\addlinespace[2pt]
\textbf{GPU counts} & 1, 2, 4, 8, 16, 32, 64, 72, 100 &
\textbf{Parallelism} & Tensor (TP) 1/2/4/8, pipeline (PP) 1/2/4/8/16, $\lfloor N/(\mathrm{TP}\times\mathrm{PP})\rfloor$ replicas for $N$ GPUs \\
\addlinespace[2pt]
\textbf{Workloads} & Short chat, long RAG, paused sessions, tool sessions, offline jobs &
\textbf{Serving} & Batch cap 1/4/16, prefill chunks of 512 or 16,384 tokens. Reuse prefixes in long RAG and paused sessions. \\
\addlinespace[2pt]
\textbf{Service targets} & Responsive, balanced, relaxed, and offline (Table~\ref{tab:slos}) &
\textbf{Load} & 4 closed-loop users total or 1/4/16 per GPU \\
\bottomrule
\end{tabularx}
\vspace{-1.5em}
\end{table*}

\section{Architecture and Evaluation}
\textbf{Memory organizations and baselines.}
HBF introduces a hardware tradeoff: \emph{shared-site} HBF replaces two or four of eight near-memory attachment sites, sacrificing near-memory capacity and bandwidth; \emph{cascaded} HBF adds flash while preserving both. We compare these layouts with conventional GPUs. RTX~4090-scale GDDR offers 24~GiB/1.008~TB/s; H200-scale HBM offers 141~GiB/4.8~TB/s. Each HBF stack holds 512~GiB per OCP Table~3~\cite{ocp}. We impose a 2~TB/s per-GPU link ceiling and 20~W per stack (2~W static, 8~pJ/read bit), yielding a 281.25~GB/s sustained read cap per stack, or 1.125~TB/s for four. Physical writes are capped at 8~GB/s/stack with 1.3$\times$ amplification.

Conventional paths stage weights through host DRAM or SSD; an optimistic MoE CPU-expert control charges movement but zero expert arithmetic. Each eight-slot domain has 256~GiB (RTX~4090) or 4~TiB (H200) host DRAM and eight SSDs, excluded from acquisition costs. Controls use static expert placement and near active KV. FlexGen can amortize reads over more microbatches than our search permits.

\textbf{Traffic and simulation.}
Our discrete-event simulator models contended GPU, HBF, host, SSD, and pipeline resources with bounded staging. Each stage reserves runtime, activations, and active KV; shared weights are pinned before experts occupy residual near memory. Cold weights stream on demand. Large MoEs use uniform routing marginals; Qwen3-30B uses published training-window frequencies~\cite{suram}. For independent token routes, the probability batch size $b$ requests cold expert $e$ is $1-(1-q_e)^b$. Thus, total weights determine whether a model fits, but the experts touched by each batch determine whether it runs fast enough.

For long RAG and paused sessions, uncached reusable blocks are written to the far cache after request completion when space is allocated; blocks become reusable after writes finish. Prefix restores charge reads and near copies. Muse sliding layers read at most 2,048 tokens; prefix-enabled designs retain all-layer archives through completion for restoration at any retained boundary.

GPU times use pinned NVIDIA AISimulate tables with an RTX~4090 timing transfer~\cite{aisim}. 
We model eight-GPU nodes with effective pipeline-transfer bandwidths of 315 GB/s for H200 and 8.4 GB/s for RTX 4090-scale GPUs, each with 8 µs startup, and sequential inter-node endpoint service at 35 GB/s plus 8 µs per endpoint, while tensor-parallel collective costs use AISimulate’s H200 predictions and the same timing transfer applied to RTX 4090-scale operations.
HBF read times are fitted to reduced-geometry HBFSim runs at queue depth one for 1- and 16-MiB ranges, subject to link/power caps~\cite{hbfsim}. Frontier and LLMServingSim provide attention/host and conventional-path checks~\cite{frontier,servingsim}. Muse uses a constructed full/sliding-attention, compressed-expert graph with transferred unmeasured timings. The resulting deployment-level results are simulation projections, not full-system measurements.

\textbf{Workloads and passing criteria.}
We use logical BF16 text weights and KV, including BF16 scenarios for FP8 releases. Workloads are short chat (512 input/64 output tokens), long RAG (16,384/64, 8,192-token shared prefix), paused sessions (4,096/64, 60-s revisit), TraceLab-derived tool sessions, and offline jobs (2,048/128)~\cite{tracelab}. Table~\ref{tab:slos} specifies targets. TTFT includes queueing; mean and p99 ITL use consecutive output-token gaps (linearly interpolated p99). At least 95\% of offered interactive requests must finish within all applicable limits per window; one-token responses use TTFT alone. Offline jobs must finish within one hour of arrival. Designs must pass the above criteria in two random windows (seeds 17, 29); useful throughput is the lower window rate, counting only passing-request tokens over total elapsed time including drain.

We enumerate Table~\ref{tab:matrix}, screen capacity and checked service bounds, then replay eligible designs. Interactive replays are scored against all three SLOs; fixed-demand screening adds demand-aware candidates. Fixed-count tool comparisons use matched node templates of replica width at most eight.

\begin{table}[!t]
\caption{Service-level objectives in the main deployment sweep}
\label{tab:slos}
\centering
\footnotesize
\begin{tabular*}{\columnwidth}{@{\extracolsep{\fill}}lccc@{}}
\toprule
\textbf{Target} & \textbf{TTFT} & \textbf{Mean ITL} & \textbf{p99 ITL} \\
\midrule
Responsive & $\leq 2$ s & $\leq 50$ ms & $\leq 100$ ms \\
Balanced & $\leq 5$ s & $\leq 100$ ms & $\leq 200$ ms \\
Relaxed & $\leq 10$ s & $\leq 200$ ms & $\leq 400$ ms \\
Offline & \multicolumn{3}{c@{}}{Completion within 1 hour} \\
\bottomrule
\end{tabular*}
\vspace{-2em}
\end{table}

\textbf{Acquisition cost.}
Fewer GPUs do not necessarily mean a cheaper system: the avoided GPU cost must exceed the HBF premium. Let $N_B$ be the minimum passing conventional GPU count. A passing HBF design $d$ uses $N_d$ GPUs and $S_d$ stacks, costing $N_d+pS_d$ GPU equivalents for all-in stack premium $p$. Its break-even ceiling, reselecting $d$ at each price, is
\begin{equation}
p^*=\max_d\frac{N_B-N_d}{S_d}.
\label{eq:premium}
\end{equation}
The premium covers flash, controller, link, and integration; shared-site prices are relative to the conventional package. Without a passing conventional design, $p^*$ is undefined. Frontiers apply to the screened pool.

\begin{figure}[t]
\centering
\includegraphics[width=\columnwidth]{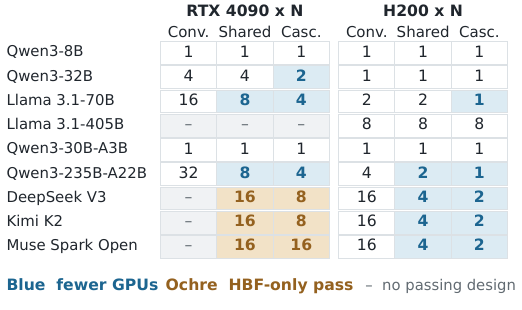}
\caption{Minimum installed GPUs for four balanced short-chat users. The conventional column includes the optimistic CPU-expert control.}
\label{fig:minimum}
\vspace{-1em}
\end{figure}

\begin{figure}[t]
\centering
\includegraphics[width=\columnwidth]{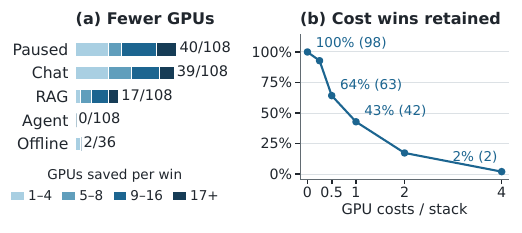}
\caption{Fixed-demand comparisons: (a) GPU-count wins by workload and GPUs saved; (b) share of 98 zero-premium wins retained with design re-selection at each stack price (counts in parentheses).}
\label{fig:costs}
\vspace{-1em}
\end{figure}

\section{Results}
We first ask whether HBF can replace GPUs without violating the SLO, then whether those savings pay for flash. Our sweep covers 810 model/platform/workload/count cells, 62,279 replays, and 17,375 service-bound exclusions. The 468 fixed-demand comparisons match each HBF organization against conventional serving under the same model, platform, workload, and SLO.

\begin{takeaway}
For multi-user serving, HBF can replace GPUs needed for weight capacity; cascaded HBF usually saves more GPUs than shared-site HBF (Fig.~\ref{fig:minimum}).
\end{takeaway}

HBF lowers the minimum GPU count in 98 of the 224 comparisons where both sides pass and conventional serving needs more than one GPU; in 82 of these wins, at most four HBF-equipped GPUs suffice. Both HBF organizations pass in 178 cases: 123 ties, 54 cascaded wins, and one shared-site win. In 29 additional HBF-versus-conventional comparisons, only HBF passes, covering 16 distinct DeepSeek V3, Kimi K2, or Muse model/workload/target combinations on RTX~4090 scale; no conventional cost reference exists for these cases.

The reductions can be large (Fig.~\ref{fig:minimum}). For four users with balanced short-chat SLOs, Qwen3-235B falls from 32 conventional RTX~4090-scale GPUs to eight with shared-site HBF or four with cascaded HBF. On H200 scale, DeepSeek V3, Kimi K2, and Muse each fall from 16 conventional GPUs to four shared-site or two cascaded. Cascaded reduces GPU count in eight of 14 cases with a conventional pass and beats shared-site in ten of 17 joint passes (seven ties). An exception is balanced long RAG with Muse on H200 scale: shared-site needs eight GPUs, cascaded 16, conventional 32.

\begin{figure}[t]
\centering
\includegraphics[width=\columnwidth]{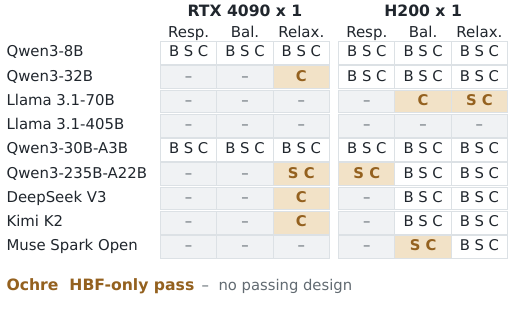}
\caption{Passing layouts for one short-chat user on one GPU. B denotes conventional, S denotes shared-site HBF, and C denotes cascaded HBF.}
\label{fig:single}
\vspace{-1em}
\end{figure}

\begin{takeaway}
HBF makes previously infeasible models serveable on a single GPU, including large MoEs on RTX~4090 scale; it also expands the SLOs met on H200 scale (Fig.~\ref{fig:single}).
\end{takeaway}

A single GPU is the clearest test of whether HBF enables a model that conventional memory cannot serve (Fig.~\ref{fig:single}). For one short-chat user, RTX~4090-scale relaxed-SLO coverage grows from two to six models: HBF brings Qwen3-32B, Qwen3-235B, DeepSeek V3, and Kimi K2 within reach. H200-scale balanced coverage rises from six to eight through Llama 3.1-70B and Muse; Qwen3-235B also gains a responsive pass. Muse has 1.49~TiB of logical BF16 text weights~\cite{muse}, exceeding two stacks plus near memory on either platform. Four stacks pass capacity screening and deliver 11.3 useful output tokens/s on H200 scale, yet Muse meets no tested single-user chat target on RTX~4090 scale. Simply fitting the weights, however, does not guarantee acceptable token latency.

\begin{takeaway}
HBF saves GPUs most often for paused sessions and short chat; benefits are smaller for long RAG and nearly absent for tool sessions and offline jobs (Fig.~\ref{fig:costs}a).
\end{takeaway}

\begin{figure}[t]
\centering
\includegraphics[width=\columnwidth]{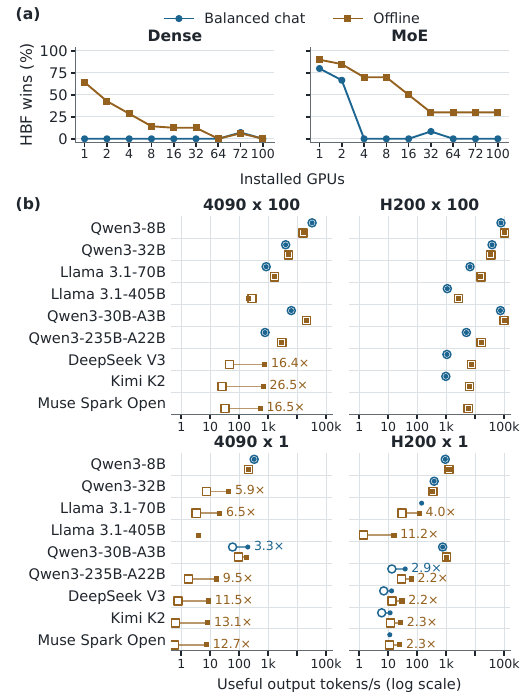}
\caption{(a) HBF win rate by model family/GPU count among joint passes. (b) Balanced chat (blue circles) and offline (ochre squares), at 1 and 100 GPUs: open conventional, filled cascaded HBF; overlaid ties, absent no pass. Labels show gains $\geq2\times$. Best passing loads.}
\label{fig:scale}
\vspace{-1em}
\end{figure}

Workload changes the answer (Fig.~\ref{fig:costs}a). HBF saves GPUs in 40/108 paused-session and 39/108 short-chat comparisons, but only 17/108 long-RAG comparisons. Of the 79 paused/chat wins, 39 save at least nine GPUs. Tool sessions yield 0/108 wins. Both offline wins (2/36) involve RTX~4090-scale Llama 3.1-405B, dropping from two GPUs to one. Why so few offline GPU-count wins? For RTX~4090-scale Kimi K2, even the conventional layout finishes four offline jobs on one GPU within the hour: HBF may finish more jobs, but cannot reduce the minimum below one GPU (Fig.~\ref{fig:scale}b).

\begin{takeaway}
Many GPU-count reductions still lower acquisition cost when an HBF stack costs half as much as a GPU (Fig.~\ref{fig:costs}b).
\end{takeaway}

A GPU-count win becomes a cost win only when the avoided GPUs pay for flash. After reselecting the cheapest passing layout at each stack price (Fig.~\ref{fig:costs}b), 63 of 98 zero-premium wins survive at half a GPU cost per stack (64\%), 42 at one GPU, and two at four GPUs. For example, serving Muse with balanced short-chat latency on H200 scale costs 16 conventional GPUs, versus two cascaded GPUs and eight stacks---six GPU equivalents if each stack costs half a GPU. This design breaks even at 1.75 GPU equivalents per stack, including the cost of integration.

\begin{takeaway}
At fixed GPU count, HBF improves throughput most often in installations below 16 GPUs and for large MoEs, although large offline jobs can also benefit at 100 GPUs (Fig.~\ref{fig:scale}).
\end{takeaway}

More throughput is a different benefit from avoiding GPUs (Fig.~\ref{fig:scale}a). At a fixed GPU count, HBF wins 96/200 matched comparisons below 16 GPUs (48\%), but only 41/304 at 16 or more (13\%). The sweep pools balanced chat/offline and selects each layout's best passing load among 1/4/16 users per GPU. MoEs win 110/262 comparisons (42\%), versus 27/242 dense (11\%). At one GPU, HBF wins 8/16 chat and 27/34 offline comparisons. At 100 GPUs, balanced chat has 16 ties and 9 losses; 6/36 offline matches win, all for RTX~4090-scale Kimi K2, Muse, and DeepSeek V3 under both HBF layouts.

The largest throughput improvements is where conventional weight offload remains costly (Fig.~\ref{fig:scale}b). On one RTX~4090-scale GPU, Qwen3-8B ties, while cascaded HBF gives 9.5$\times$ for Qwen3-235B and 11.5--13.1$\times$ for the three largest MoEs. Llama 3.1-405B gains 11.2$\times$ on one H200-scale GPU. On 100 RTX~4090-scale GPUs, cascaded HBF raises offline throughput from 25.7 to 681.1 tokens/s for Kimi K2 (26.5$\times$), 32.6 to 536.3 for Muse (16.5$\times$), and 47.0 to 770.8 for DeepSeek V3 (16.4$\times$). Yet the benefit is not universal: at 100 H200-scale GPUs, conventional and cascaded Muse tie at 5,774 tokens/s.

\begin{figure}[t]
\centering
\includegraphics[width=\columnwidth]{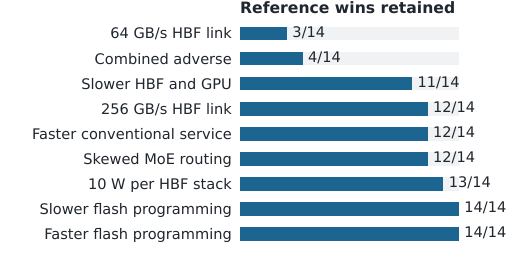}
\caption{Sensitivity of 14 GPU-count wins after reselecting designs within one replay window; surviving HBF designs still beat the conventional minimum.}
\label{fig:sensitivity}
\vspace{-1em}
\end{figure}

\section{Conclusion and Limitations}
HBF is most valuable when weights, rather than computation, force a larger GPU installation---and only when flash can deliver them within the SLO at a price below the avoided GPUs. Paused sessions and short chat show the clearest GPU-count savings in our sweep. For large MoEs, HBF can also enable single-GPU serving or greatly raise offline throughput, even when the minimum GPU count does not change.

Results depend on the modeled hardware and traffic. In a sensitivity audit of six model/platform/workload cases across three SLOs and two layouts, only two of 14 reference GPU-count wins survive all nine settings (Fig.~\ref{fig:sensitivity}); a 64-GB/s link preserves three and the combined adverse setting four. That setting limits links to 256~GB/s, slows HBF/GPU service, raises cold-weight overhead, and speeds conventional serving. Muse timing transfers have 15.4\% and 32.3\% median absolute percentage errors on neighboring-expert and held-out sliding-decode checks, respectively. For Qwen3-30B, held-out current-batch routing increases positive weight traffic by 6.7--16.4\%, without changing balanced feasibility across 60 designs (public Q4 routes, BF16 payloads). BF16 scenarios, static placement, screened layouts, and RTX~4090-scale multi-GPU assumptions limit generalization; quantization, speculative decoding, and alternative host or expert-parallel policies may change the conclusions.

\textbf{Artifact.}
Code, input provenance, exact replay receipts, and full results for both sweeps are available at \url{https://github.com/ryeii/When-Is-HBF-Cost-Effective}.
\vspace{-1em}
\newcommand{\papertitle}[1]{\textquotedblleft#1\textquotedblright}

\end{document}